# Gender-specific Call of Duty: A Note on the Neglect of Conscription in Gender Equality Indices

Jussi Heikkilä & Ina Laukkanen

ARTICLE

# Gender-specific Call of Duty: A Note on the Neglect of Conscription in Gender Equality Indices

Jussi Heikkilä [a] and Ina Laukkanen[b,c]

[a]Jyväskylä University School of Business and Economics, University of Jyväskylä, Jyväskylä, Finland; [b]Department of Social Sciences and Philosophy, University of Jyväskylä, Jyväskylä, Finland; [c]Faculty of Social Sciences, University of Helsinki

**ABSTRACT**

We document that existing gender equality indices do not account for gender-specific mandatory peace-time conscription (compulsory military service). This suggests that gender-specific conscription is not considered to be an important gender issue. If an indicator measuring the gender equality of mandatory conscription was to be included in gender equality indices with appropriate weight, then the relative rankings of countries in terms of measured gender equality could be affected. In the context of the Nordic countries, this would mean that Finland and Denmark – the countries with mandatory conscription for men only – would have worse scores with respect to gender equality compared to Sweden and Norway, countries with conscription for both men and women – and Iceland, which has no mandatory conscription, regardless of gender.

## Introduction

According to empirical evidence, there exists a positive association between women's empowerment and economic development (Duflo 2012) and gender[1] equality policies and legislation are adopted in the majority of countries around the world (Ertan 2016). Gender equality is among the United Nations (UN) Sustainable Development Goals (SDGs) and the aim is to 'end all forms of discrimination against all women and girls everywhere'.[2] Goal 5 of the SDGs is to 'achieve gender equality and empower all women and girls' and it also includes multiple subtargets. The Charter of Fundamental Rights of the European Union (EU; Article 23) states: 'Equality between women and men must be ensured in all areas, including employment, work and pay'.

A variety of gender equality indices measure the relative progress in gender equality across countries (Permanyer 2010; Bericat 2012; Hawken & Munck 2013; Ertan 2016). Multiple international organizations, such as the UN and its suborganizations (e.g., the United Nations Economic Commission for Europe, UNECE,[3] the United Nations Educational, Scientific and Cultural Organisation, UNESCO[4] and the United Nations Development Program, UNDP), the World Bank, World Economic Forum, the Organisation for Economic Co-operation and Development (OECD)[5] and the European Institute for Gender Equality (EIGE),[6] systematically track the development of gender equality across countries. Generally, gender equality has progressed in several fronts including military service.[7] Despite the general trend in shifting from conscription-based military service to professional armies, some countries still have conscription systems (Joenniemi 2006; Keller, Poutvaara, and Wagener 2009; Torun 2019). Since the beginning of the 1990s, 17 of 28 EU countries have abolished conscription; as of 2017, seven still have them in place (Torun 2019). Recently, Norway and Sweden became the first countries in the world to conscript men and women on equal terms: in 2015 and 2018, respectively (Persson and Sundevall 2019). While there exists plenty of sex-disaggregated statistics for gender pay gap, gendered labour markets, gendered violence, etc., to our knowledge, sex-disaggregated time-series data on military service is not available.[8] A lack of systematic data means that military service remains a field in which the gendered nature is to a large extent neglected.

Several studies suggest that the empowerment of women, higher status of women and gender equality are positively associated with sustaining peace and avoiding violent conflicts (e.g., Caprioli 2000, 2005; Caprioli and Boyer 2001; Melander 2005; Gizelis 2009; Crespo-Sancho 2017). The UN Security Council's Resolution 1325 – the first resolution on Women, Peace and Security – was adopted on 31 October 2000[9] and acknowledges „the important role of women in the prevention and resolution of conflicts and in peace-building, and stressing the importance of their equal participation and full involvement in all efforts for the maintenance and promotion of peace and security, and the need to increase their role in decision-making with regard to conflict prevention and resolution". Since 2000, the number of female peacekeepers has increased (Karim and Beardsley 2013; UN Peacekeeping 2019). Generally, increased attention is allocated to gender issues and the role of women in military (NATO 2017; Reis and Menezes 2019).

The contribution of this article is twofold. First, we shed light on the treatment of gender-specific mandatory conscription in existing gender equality indices. In other words, we answer the simple research question: Do gender equality indices account for gender-specific conscription? Second, we analyse the association between gender equality indices and conscription systems in the Nordic countries. The Nordic countries provide a particularly interesting case since they are among the most gender-equal countries in the world according to several gender equality indices (OECD 2018; WEF 2019) but have significant variations in their conscription systems.

This paper is structured as follows: Section 2 reviews the literature on gender and conscription; Section 3 describes gender equality indices and the role of gender-specific conscription therein; Section 4 analyses the status of gender equality and the conscription systems in the Nordic countries; and Section 5 discusses our findings and their implications.

## Gender and Conscription

According to economics textbooks, National defence is a common example of a public good in economics textbooks (e.g., Mankiw 2018, p.214). While both women and men pay taxes from which military expenditures are covered, over time this public good has been mainly produced by young men who constitute majority of soldiers.[10] The duration of a typical military service varies between six and 12 months, but in some countries it can even be multiple years.[11] Male-specific conscription systems are based on gender-discriminatory laws (cf. Roy 2019). Depending on the perspective, compulsory military service can be regarded as a tax[12] on or an investment in an individual conscript. Irrespective of the interpretation, the conscription institution has mainly affected and continues to predominantly affect male citizens directly and female citizens indirectly.

Extensive empirical evidence indicates that conscription may have a variety of impacts on people, including labour market outcomes and earnings (Card and Cardoso 2012; Galiani, Rossi, and Schargrodsky 2011; Torun 2019), education paths (Sharp and Krasnesor 1968; Keller, Poutvaara, and Wagener 2010; Di Pietro 2013; Lyk-Jensen 2018), family formation (Keller, Poutvaara, and Wagener 2010), belief and value formation (Ertola Navajas et al. 2019) and criminal behaviour (Galiani, Rossi, and Schargrodsky 2011; Lyk-Jensen 2018; Hjalmarsson and Lindquist 2019). If conscription is gender-specific, then only the chosen gender is subject to the majority of direct impacts. Therefore, conscription is also a gender issue.[13] Ignorance about this topic signals that gender equality and inclusiveness in conscriptions are not considered to be goals worth promoting.

While mandatory conscription is gender-based and male-specific in many countries, it may have indirect impacts on women as well. Torun (2019) studies the effect of peacetime conscription on the early labour market outcomes of potential conscripts before they are called up for service. It was shown that the abolition of conscription in Spain increased labour force participation and employment and decreased unemployment of teenage men, whereas the opposite effects were found for teenage women who were not subject to conscription. These findings suggest a high degree of substitutability between young men and women in the labour market.

Gender and military is a relatively little but increasingly studied topic in the literature (see Segal 2006; Reis and Menezes 2019). Military remains a gendered institution and a stereotypical soldier is considered to be male and a masculine role (Kwon 2000; Boldry, Wood, and Kashy 2001; deGroot 2001; Kronsell and Svedberg 2001; Sasson-Levy 2003; Lahelma 2005). In the 2000s, Western militaries have underwent significant changes and there has been a general trend of transition from mandatory conscription to professional militaries (Joenniemi 2006; Keller, Poutvaara, and Wagener 2009; Torun 2019) and, concurrently, women's participation in militaries has increased (Karazi-Presler, Sasson-Levy, and Lomsky-Feder 2018). According to Karazi-Presler, Sasson-Levy, and Lomsky-Feder (2018), several important supranational legal decisions, including the UN Resolution 1325 and the North Atlantic Treaty Organization's (NATO) adoption of a gender mainstreaming approach, have led to increased participation of women in Western militaries. A recent systematic review by Reis and Menezes (2019) of the literature related to 'women' and 'army' suggests that the majority of studies were conducted by US-based military sociologists and focused on topics such as 'sexual assault and harassment in active-duty military', 'femininity and egalitarianism' and 'posttraumatic stress disorder'.

Despite an increase in attention that is allocated to the gender analysis of military and conscription in various countries (e.g., South Korea: Kwon 2000; Israel: Sasson-Levy 2003; Sasson-Levy 2011; Sweden: Persson and Sundevall 2019), surprisingly little attention has been allocated to quantifying gender (in)equality with respect to conscription. To our knowledge, none of the existing gender equality indices include gender-specific mandatory conscription. Next, we will systematically review these indices.

## Gender Equality Indices

There exists a variety of gender equality indices that aim to measure the status and progress of countries with respect to gender equality (Sugarman and Straus 1988; Di Noia 2002; Schüler 2006; Permanyer 2010; Bericat 2012; Hawken & Munck 2013; Bericat and Sánchez Bermejo 2016; Dilli, Carmichael, and Rijpma 2019; World Bank 2019). Multiple authors have outlined good practices in measuring gender equality. According to Dijkstra (2006, p.276), '(1) It should cover a limited number of indicators, but these indicators together should cover as many dimensions of gender equality as possible; (2) Data should be available for many countries; (3) It should be simple to calculate and to understand; (4) It should allow comparisons between countries but also over time'. Bericat and Sánchez Bermejo (2016) note: 'Given that gender equality is a multidimensional phenomenon, its measurement requires the help of a composite indicator, in other words, a series of individual indicators compiled into a single index based on an underlying model (Nardo et al. 2008:13)'. According to Plantenga et al. (2009), 'A useful index should serve three main goals: to identify the extent of gender (in)equality at a certain point in time; to identify causes for (in)equality with a view to suggesting policies to reduce inequality; and finally, to enable the monitoring of the impact of these policies over time'.

Table 1 provides a non-comprehensive list of gender equality indices. We focused on selected indices that are applicable for country comparisons. None of these indices takes into account gender (in)equality in conscription. This suggests that gender-specific conscription is not considered to be a gender equality issue of high importance. The neglect of including conscription in the indices implies that reforms on gender-specific conscription do not affect gender equality when it is measured using the indices listed in Table 1.

Table 1. Selected gender equality indices.

| Index | Abbreviation | Organization or author | Since | Description | Gender-specificity of conscription acknowledged |
|---|---|---|---|---|---|
| Gender-related Development Index | GDI | UNDP | 1995 | The GDI measures gender gaps in human development achievements by accounting for disparities between women and men in three basic dimensions of human development – health, knowledge and living standards using the same component indicators as in the HDI. The GDI is the ratio of the HDIs calculated separately for women and men using the same methodology as in the HDI. It is a direct measure of gender gap showing the female HDI as a percentage of the male HDI. | No |
| Gender Empowerment Measure | GEM | UNDP | 1995 | The GEM examines whether women and men are able to actively participate in economic and political life and take part in decision-making. | No |
| Relative Status of Women | RSW | Dijkstra and Hanmer (2000) | 2000 | The RSW is based on the same indicators as the GDI (and the HDI): educational attainment, longevity, and income. However, it is a relative measure that assesses the position of women compared to that of men. | No |
| African Gender Status Index | AGSI | UNECA | 2004 | The AGSI consists of three blocks: the social power, which measures human capabilities; the economic power, which measures economic opportunities; and the political power, which measures voice or political agency. | No |
| European Union Gender Equality Index | - | EIGE | 2005 | A composite indicator that measures the complex concept of gender equality and, based on the EU policy framework, assists in monitoring progress of gender equality across the EU over time. | No |
| Global Gender Gap Index | GGGI | WEF | 2006 | A framework for capturing the magnitude of gender-based disparities and tracking their progress over time. The Index benchmarks national gender gaps on economic, education, health and political criteria, and provides country rankings that allow for effective comparisons across regions and income groups. The rankings are designed to create global awareness of the challenges posed by gender gaps and the opportunities created by reducing them. | No |
| Gender Equity Index | GEI | Social Watch | 2007 | The GEI is based on information available that can be compared internationally, and it makes it possible to classify countries and rank them in accordance with a selection of gender inequity indicators in three dimensions, education, economic participation and empowerment. | No |
| Women, Business and the Law | WBL | The World Bank | 2008 | The indicator measures how laws affect women throughout their working lives. | No |
| Social Institutions and Gender Index | SIGI | OECD | 2009 | Composite measure of gender equality, based on the OECD's Gender, Institutions and Development Database. | No |
| Gender Inequality index | GII | UNDP | 2010 | GII measures the human development costs of gender inequality. The higher the GII value the more disparities between men and women and the more loss to human development. | No |
| SDG Gender Index | - | Equal Measures 2030 | 2018 | The 2019 SDG Gender Index measures the state of gender equality aligned to 14 of the 17 Sustainable Development Goals (SDGs) in 129 countries and 51 issues ranging from health, gender-based violence, climate change, decent work and others. | No |
| Historical Gender Equality Index | HGEI | Dilli, Carmichael, and Rijpma (2019) | 2019 | HGEI measures gender equality in four dimensions: socioeconomic, health, household, and politics. | No |
| Basic Indicator of Gender Inequality | BIGI | Stoet & Geary (2019) | 2019 | Basic Index of Gender Inequality (BIGI) is the ratio of women to men on three core dimensions of life; 1) Educational opportunities in childhood; 2) Healthy life expectancy the number of years one can expect to live in good health); and, 3) Overall life satisfaction. | No |

In addition to the indices presented in Table 1, there exists a variety of other gender equality indices that measure gender equality within specific countries or between states of specific countries. These include, for instance, Yllö (1984), Sugarman and Straus (1988) and Di Noia (2002) for the U.S. states, Harvey, Blakely, and Tepperman (1990) for Canada and Frias (2008) for Mexican states. These indices also do not consider gender equality in conscription.

## Gender Equality and Gender-specific Conscription: The Case of the Nordic Countries

### Gender Equality

As noted, the Nordic countries have consistently been ranked among the most gender-equal countries in the world when gender equality is measured using existing gender equality indices. Figure 1 shows the trends of selected indices over time. The indices are not consistent with respect to the gender-equality rankings of countries, but they clearly show that differences between the Nordic countries are relatively small. Notice that there have been substantial changes in the conscription systems of Norway and Sweden: Norway introduced equal conscription for men and women in 2015, while Sweden abolished male-specific conscription in 2010 and introduced conscription for both men and women in 2017.

Each of the Nordic countries occasionally publishes reports on the development of gender equality. We reviewed the most recent versions of these reports to determine whether they include statistics on gender equality with respect to military service. Since Iceland does not have an army, it was not considered in this analysis. We found that none of the publications report any statistics on gender equality in terms of military service (Statistics Finland 2018; Statistics Norway 2018; Statistics Sweden 2018).[14] However, Statistics Finland (2018) mentions as 'a milestone of gender equality' year 1995 when 'voluntary military service became a possibility for women' and Statistics Sweden (2018) notes that in 2010, 'A change in the National Total Defence Act makes conscription gender neutral'. Furthermore, Norden (2015) report on gender equality in the Nordic countries also does not consider gender equality in military service.

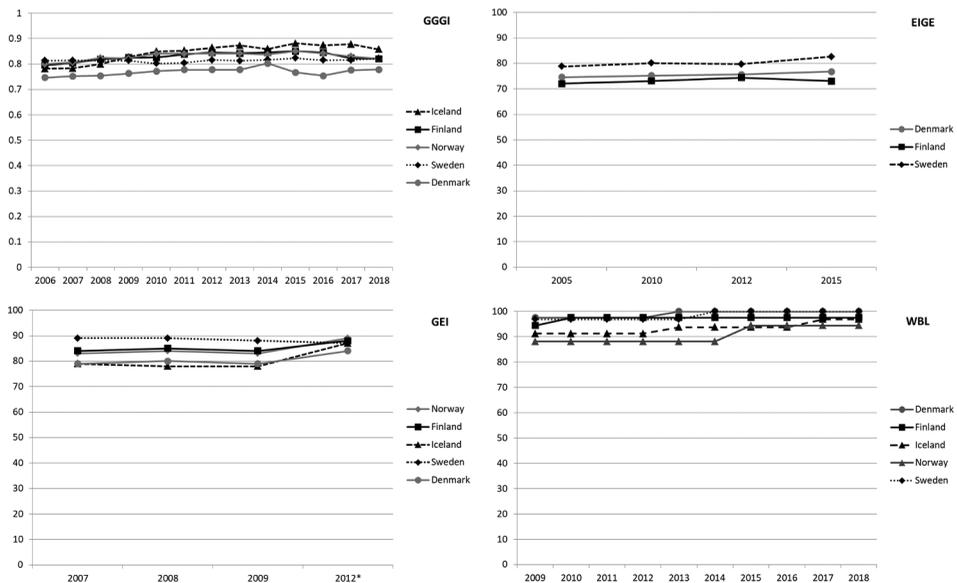

**Figure 1.** Development of selected gender equality indices over time. Notes: Authors' illustrations. See Table 1 for descriptions of the indices. Information sources for gender equality index values are WEF (2019) for Global Gender Gap Index, EIGE (2017) for EIGE Gender Equality Index, Social Watch (2012) for Gender Equity Index (GEI) and World Bank (2019) for Women, Business and the Law Index (WBL).

## Conscription Systems

In this section, we briefly describe the national conscription systems and country specificities in legislation related to gender equality.

### Norway

According to the Norwegian Equality and Anti-discrimination Act, its purpose is to 'promote equality and prevent discrimination on the basis of gender, ... gender identity, gender expression, age or other significant characteristics of a person'.[15] Norway introduced general conscription for men in 1897. In 2015, Norway became the first country in the world to conscript men and women on equal terms. According to the Defence Act, „Norwegian nationals who are fit for service in the Armed Forces have a military duty from the year they turn 19, to the end of the year they turn 44 years" but „the duty of protection does not apply to women born before 1 January 1997". In 2018, 25.2% of conscripts were women (see Table 2). Norway is one of the founding members of NATO.

### Sweden

According to the Swedish Discrimination Act, its purpose is to „combat discrimination and in other ways promote equal rights and opportunities regardless of sex, transgender identity or expression, ethnicity, religion or other belief, disability, sexual orientation or age."[16] The Swedish Conscription Act states that „national defense is a matter for the entire population. Conscription applies to every Swedish citizen from the beginning of the calendar year when he or she turns sixteen to the end of the calendar year when he or she turns seventy."[17] Conscription for men was introduced in Sweden in 1901 and remained in place until 2010 (Joenniemi 2006). According to Persson and Sundevall (2019), gender-neutral military conscription was raised for public debate for the first time in Sweden in the mid-1960s, and male-specific conscription was abolished in 2010. It was replaced with a gender-neutral conscription that was, however, inactive during peace time. Conscription was reactivated in 2017 in its gender-neutral form and the first female conscripts started in 2018. Sweden became the second country in the world, after Norway, to conscript men and women on equal terms (Persson and Sundevall 2019). In 2018, 16% of new conscripts were women (see Table 2). Sweden is not a member of NATO.

### Denmark

According to the Danish Equality Act for women and men,[18] its purpose is to "promote equality between women and men, including equal integration, equal influence and equal opportunities in all functions of society based on the equal value of women and men. The purpose of the Act is also to discourage direct and indirect discrimination on the grounds of sex ... ". According to the Constitution of Denmark, which was founded in 1849, all physically fit men are obligated to conscript and the Danish Military Service Act states that „Every Danish man is subject to military service."[19] However, military service is voluntary for women. In 2018, 16.8% of conscripts were women (see Table 2). Denmark is a founding member of NATO.

### Iceland

According to the Iceland Act on the Equal Status and Equal Rights of Women and Men, its aim is to 'establish and maintain equal status and equal opportunities for women and men, and thus promote gender equality in all spheres of the society. All individuals shall have equal opportunities to benefit from their own enterprise and to develop their skills irrespective of gender. This aim shall be reached

Table 2. Selected gender equality indices and conscription systems.

| | Finland | Sweden | Norway | Denmark | Iceland |
|---|---|---|---|---|---|
| Gender equality index value (ranking) | | | | | |
| Global Gender Gap Index (GGGI) in 2018 | 0.821 (#4) | 0.822 (#3) | 0.835 (#2) | 0.778 (#13) | 0.858 (#1) |
| EIGE Gender Equality Index in 2015 | 73.0 (#3) | 82.6 (#1) | - | 76.8 (#2) | - |
| Gender Inequality Index (GII) in 2017 | 0.058 (#8) | 0.044 (#3) | 0.048 (#5) | 0.04 (#2) | 0.062 (#9) |
| Social Institutions and Gender Index in 2019 | 15.3% (#15) | 10.5% (#3) | 14.5% (#12) | 10.4% (#3) | - |
| Women, Business and the Law (WBL) Index in 2018 | 97.5 (#2) | 100 (#1) | 94.38 (#5) | 100 (#1) | 96.88 (#3) |
| Conscription in 2018 | | | | | |
| Mandatory for men | Yes | Yes | Yes | Yes | No |
| Mandatory for women | No | Yes | Yes | No | No |
| Number of conscripts in 2018 | | | | | |
| Men | 23,930 | 3,150 | 5,493 | 3,502 | 0 |
| Women | 943 | 600 | 1853 | 706 | 0 |
| Share of women | 3.8 % | 16.0 % | 25.2 % | 16.8 % | - |
| Population in 2018* | 5,518,050 | 10,183,170 | 5,314,340 | 5,797,450 | 353,570 |
| Military expenditures in 2018 (% of govt. spending)** | $3,949 M (2.65%) | $5,755 M (2.15%) | $7,067 M (3.41%) | $4,228 M (2.31%) | - |
| World War II casualties (estimate)*** | 85,000–97,000 | Hundreds | ca. 10,000 | Thousands | - |
| NATO member**** | No | No | Yes | Yes | Yes |

Notes: Information sources for gender equality index values and rankings are WEF (2019) for Global Gender Gap Index, EIGE (2017) for EIGE Gender Equality Index, UNDP (2018) for GII, OECD (2019) for Social Institutions and Gender Index and World Bank (2019) for Women, Business and the Law Index. The number of conscripts data is collected from webpages of armed forces for Norway, Sweden and Denmark and from Statistics Finland. In the case of Finland, Denmark and Sweden, the numbers refer to the number of citizens starting the military service for Finland, Denmark and Sweden and in the case of Norway to the number of citizens that have completed military service. Iceland is not included as there is no conscription. The data is available from the authors upon request. * Source: World Bank, https://data.worldbank.org/indicator/sp.pop.totl ** Current US dollars. Source: Stockholm International Peace Research Institute, SIPRI Military Expenditure Database, https://www.sipri.org/databases/milex *** Source: Tham (1990) **** Source: https://www.nato.int/cps/en/natohq/nato_countries.htm

by: a. gender mainstreaming in all spheres of the society, b. working on the equal influence of women and men in decision-making and policy-making in the society, c. enabling both women and men to reconcile their occupational and family obligations, d. improving especially the status of women and increasing their opportunities in the society, e. increasing education in matters of equality, f. analysing statistics according to sex. g. increasing research in gender studies'.[20] Iceland does not have its own army and conscription system. However, Iceland has a Coast Guard and is a founding member of NATO.

### Finland

According to the Finnish Act on Equality between Women and Men, its objectives are to "prevent discrimination based on gender, to promote equality between women and men, and thus to improve the status of women, particularly in working life. Furthermore, it is the objective of this Act to prevent discrimination based on gender identity or gender expression".[21] In Finland, mandatory conscription has been in place for men since 1878 when Finland was a grand duchy of the Russian empire with extensive autonomy. The Finnish Constitution states regarding national defence obligation (Section 127) that „every Finnish citizen is obligated to participate or assist in national defence, as provided by an Act". According to the Finnish Conscription Act, "Every male Finnish citizen is liable for military service starting from the beginning of the year in which he turns 18 years old until the end of the year in which he turns 60, unless otherwise provided for herein".[22] When the Finnish Parliament passed the Equality Act, it determined that compulsory military service for men does not constitute discrimination as prohibited by the Equality Act. Section 9 of the Act on equality between women and men (2005) lists "actions that shall not be deemed to constitute discrimination" and states that "enacting legal provisions on compulsory military service for men only" shall not be deemed to constitute discrimination based on gender. In 2018, the length of military service varied between six and 12 months and the share of the male cohort (born in 2000) that was ordered to complete compulsory military service in call-ups was about 75%. There is also a 12-months non-military service option for men that do not want to complete military service.[23] In 2018, the share of voluntary women of all conscripts was 3.8% (see Table 2). Finland is not a member of NATO.

### Summary

To summarize, the Nordic countries are consistently regarded as the most gender-equal countries in the world and the differences between the countries are relatively small according to existing gender equality indices. However, with respect to gender-specific conscription, they are a heterogeneous set. From the perspective of gender equality, at one end of the spectrum, Norway and Sweden have voluntary service for both men and women, while at the other end, Denmark and Finland have mandatory conscription for men only. Iceland does not have a conscription system and could be interpreted to be gender-equal. Table 2 provides a summary of key statistics for the Nordic countries. It also includes the estimates of the number of casualties in World War II. These figures demonstrate the disproportionate impacts across countries and provide historical context for national military systems.

Figure 2 shows the development in the number of conscripts by gender and the trends in the share of women conscripts over time. During the period, Norway increased the number and share of women in its conscripts the most. It is clear that in Finland, the conscription system is the least gender equal, as the share of women is in the low single-digits, whereas in Sweden, Denmark and Norway, the share of women beginning military service has varied approximately between 15% and 25% in recent years.

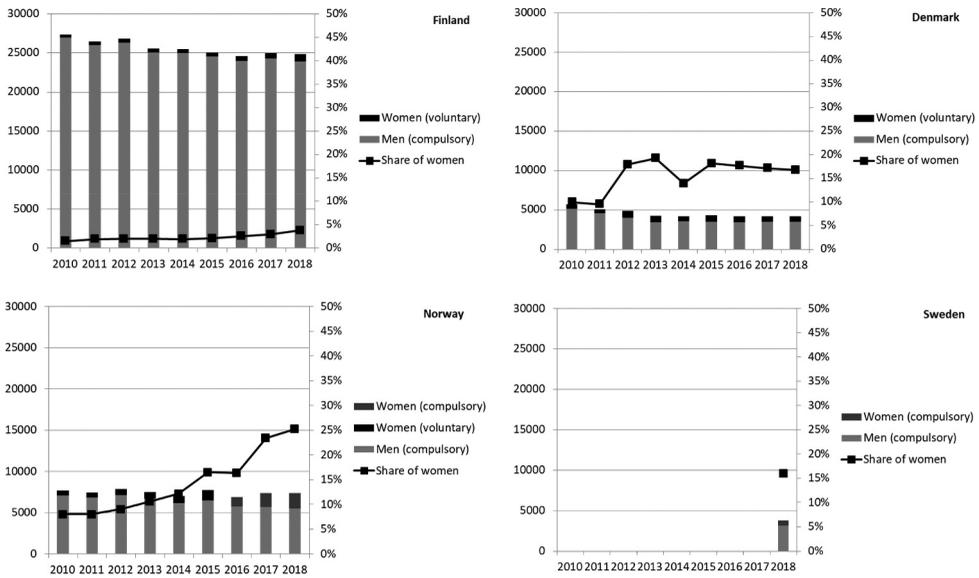

**Figure 2.** Women and men in military service (2010–2018). Notes: The data is collected from the webpages of the armed forces of Norway, Sweden and Denmark and from Statistics Finland. In the case of Finland, Denmark and Sweden, the numbers refer to the number of citizens starting the military service for these countries, and in the case of Norway, the numbers refer to the number of citizens that have completed military service. Iceland is not included, as it does not have an army or conscription. The data is available from the authors upon request.

## Discussion and Conclusions

We document that existing gender equality indices do not account for gender-specific mandatory conscription. If the indices included indicators measuring gender equality in conscription, the relative ranking of the countries with respect to gender equality could be impacted. Their impact depends naturally on the weight given to this particular section within the aggregate index. Currently, the weight is zero and it is questionable whether this is optimal when aiming to measure progress in gender equality.

As pointed out by developers of gender equality indices, useful gender equality indices should 1) identify the extent of gender (in)equality at a certain point in time, 2) identify causes for (in)equality with a view to suggesting policies to reduce inequality and 3) enable the monitoring of the impact of these policies over time (Plantenga et al. 2009). By neglecting gender-specific conscription, current indices are incomplete and 1) signal that gender-specific conscription is not considered to be an important gender equality issue and 2) suggest that reforms in gender-specific conscription do not affect measured gender equality. One potential reason for the neglect of gender-specific military service might be that the majority of countries have already abolished conscription systems altogether (Torun 2019) and, therefore, conscription is no longer a gender issue.

Different gender quality indices assign different weights to varying sets of gender equality dimensions and it is not a straightforward task to define an appropriate weight to gender-specific conscription relative to other dimensions. Nonetheless, if we assume that gender-neutral conscription (the same irrespective of gender) is more gender equal compared to gender-specific conscription, then it would be appropriate to assign better scores for the former compared to the latter. In the context of the Nordic countries, this would mean that Finland and Denmark – the countries with conscription for men only – would have worse scores with respect to gender equality compared to Sweden and Norway – the countries with conscription for men and women – and Iceland, which does not have conscription. The main implication of this research note for future studies is that they could focus on creating new gender equality indices that account for gender-specific conscription.

The neglect of military conscription may also bias other indicators of gender equality. Some gender equality indices include measures of the average years of education by gender (e.g., Bericat 2012; Dilli, Carmichael, and Rijpma 2019). However, military education is not considered as education in these statistics. If military service was considered as education, its impact on average years of schooling would be disproportional by gender. For instance, in Finland in 2017, the average years of schooling for women was 12.6 and for men 12.3 according to the UN's Gender Development Index.[24] Since military service varies between five and 12 months, including military education would diminish the gap between men and women in terms of average years of schooling. Generally, research results concerning the impact of military service on education are not conclusive (Sharp and Krasnesor 1968; Keller, Poutvaara, and Wagener 2010; Di Pietro 2013; Lyk-Jensen 2018), but it is clear that the opportunity cost of military service for several conscripts is tertiary education investment or work experience, and it is a fact that time spent in military service shortens the length of an individual's career. It is also a fact that the education level of women has continually increased and bypassed that of men in several countries (Bericat and Sánchez Bermejo 2016; WEF 2019).

In recent years, gender-responsive budgeting has been introduced in several countries, including the Nordic countries (Stotsky 2016; Downes, von Trapp, and Nicole 2016).[25] The Council of Europe (2009) defines gender budgeting as 'an application of gender mainstreaming in the budgetary process. It means a gender-based assessment of budgets, incorporating a gender perspective at all levels of the budgetary process and restructuring revenues and expenditures in order to promote gender equality'. Military investments typically constitute a non-negligible share of government budgets as shown in Table 2. When conscription is gender-specific, the government necessarily discriminates by gender in its investments on the military skills of its citizens. On the other hand, conscription could be considered to be a tax (Poutvaara and Wagener 2009), and since it is typically compulsory only for young men, it could be interpreted to be a gender-based tax (cf. Alesina, Ichino, and Karabarbounis 2011; Colombino and Narazani 2018). These gender-specific investments or taxes are presumably of interest for future studies of gender budgeting (Hendricks and Hutton 2008).

## Acknowledgements

Earlier versions of this paper were presented at the Allecon seminar and the Gender Studies Conference 2020 in Tampere. We thank Terhi Ravaska, all the participants of the seminars and two anonymous reviewers for helpful comments.

## Notes

1. We acknowledge the difference between sex and gender: „sex" refers to an individual's biological characteristics whereas „gender" refers to an individual's social identity (EIGE 2014). For simplicity and consistency, we use throughout the paper the term gender instead of sex.
2. https://www.un.org/en/sections/issues-depth/gender-equality/ and https://www.un.org/sustainabledevelopment/gender-equality/. Accessed 23 September 2019.
3. http://www.unece.org/stats/gender.html. Accessed 23 September 2019.
4. https://en.unesco.org/genderequality. Accessed 23 September 2019.
5. http://www.oecd.org/gender/. Accessed 23 September 2019.
6. https://eige.europa.eu/. Accessed 23 September 2019.
7. We use the terms „conscription" and „compulsory military service" interchangeably throughout this paper. We use the term „conscription" as a synonym for „peacetime conscription" for the sake of brevity and simplicity.
8. E.g., the Gender Data Portal of the World Bank: http://datatopics.worldbank.org/gender/. Already Evans, Felson, and Kenneth (1980) have made the observation that „social indicator research has seldom considered the social conditions of the military and their interrelationships with the social conditions of civilian society" (p.81) and note that „none of the 150 indicators of American social conditions ... is specifically concerned with the military or its impact on the larger society" (p.82).
9. https://www.usip.org/gender_peacebuilding/about_UNSCR_1325. Accessed 10 September 2019.

10. It should be noted that women have been largely present in various civilian positions producing national defence. We thank an anonymous reviewer for emphasizing this.
11. See https://www.cia.gov/library/publications/the-world-factbook/fields/333.html. Accessed 29 October 2019.
12. Poutvaara and Wagener (2009) note that „economically, a military draft is a tax in the form of coerced and typically underpaid labor services". Military draft could be also be viewed as statute labour that is 'unpaid work on public projects that is required by law' according to Encyclopedia Britannica. See https://www.britannica.com/topic/statute-labour. Accessed 22 August 2020.
13. According to the EIGE (2014, 16), „Gender issues are all aspects and concerns related to how women and men interrelate, their differences in access to and use of resources, their activities, and how they react to changes, interventions, and policies".
14. Statistics Denmark did not have a report, but we obtained information from the Statistics Denmark's Gender Equality Website instead. https://www.dst.dk/en/Statistik/emner/levevilkaar/ligestilling/ligestillingswebsite. Accessed 7 October 2019.
15. https://lovdata.no/dokument/NLE/lov/2017-06-16-51. Accessed 20 September 2019. Unofficial translation.
16. https://www.government.se/information-material/2015/09/discrimination-act-2008567/. Accessed 20 September 2019.
17. Lag (1994:1809) om totalförsvarsplikt, https://www.riksdagen.se/sv/dokument-lagar/dokument/svensk-forfattningssamling/lag-19941809-om-totalforsvarsplikt_sfs-1994-1809. Accessed 20 September 2019. Unofficial translation.
18. https://www.retsinformation.dk/Forms/R0710.aspx?id=160578. Accessed 20 September 2019. Unofficial translation.
19. https://www.retsinformation.dk/forms/r0710.aspx?id=6463. Accessed 20 September 2019. Unofficial translation.
20. http://www.humanrights.is/en/moya/page/act-on-the-equal-status-and-equal-rights-of-women-and-men. Accessed 20 September 2019. Unofficial translation.
21. https://www.finlex.fi/en/laki/kaannokset/1986/en19860609_20160915.pdf. Accessed 20 September 2019. Unofficial translation.
22. https://www.finlex.fi/fi/laki/kaannokset/2007/en20071438.pdf. Accessed 20 September 2019. Unofficial translation. The act also states that: „Fulfilment of military service includes service as a conscript, participation in reservist training, extra service, and service during mobilization, in addition to participation in call-ups and examinations assessing fitness for military service". The maximum number of reservist training days varies from 80 to 200 days conditional on the military rank of the reservist.
23. Non-Military Service Act (1446/2007), https://www.finlex.fi/fi/laki/kaannokset/2007/en20071446_20130940.pdf. Accessed 10 October 2019.
24. http://hdr.undp.org/en/composite/GDI. Accessed 21 September 2019.
25. https://www.imf.org/external/datamapper/datasets/GD. Accessed 20 October 2019.

## Disclosure Statement

No potential conflict of interest was reported by the authors.

12  J. HEIKKILÄ AND I. LAUKKANEN